\documentstyle[aps,prl,psfig]{revtex}

\begin{document}
\twocolumn
\noindent
{\bf Comment on ``Point-Contact Study of Fast and Slow Two-Level
Fluctuators in Metallic Glasses''}

In a beautiful recent experiment on  mechanically controlled
break junctions made from metallic glasses  \cite{KSK},
Keijsers, Shklyarevskii and van Kempen (KSK) 
found a zero-bias anomaly (ZBA) in the differential conductance
that switched between two  or more values (switching times $> \! 1$s). 
The $V$-dependence of the fluctuation amplitude $\Delta G(V)= |G-G'|$,
shown in Fig.~1 for two of KSK's samples, implies that this is not simply
a standard telegraph-noise-like signal superimposed on a ZBA, since
then $\Delta G$ would be constant. KSK attributed the ZBA to
{\em fast}\/ two-level
systems (TLSs) in the junction, 
and its telegraph-like fluctuations to the modulation 
of some fast TLSs' parameters, induced by short-ranged interactions with
nearby, {\em slowly}\/ switching  two-state systems.

KSK found that if a {\em distribution}\/ of  TLS parameters is assumed, 
the ZBA's overall shape is consistent
with  both the theories of Kozub and Kulik (KK) \cite{KK}
and Vlad\'ar and Zawadowski (VZ) \cite{VZ,Ralph,vD} for the TLS-electron
interaction.
In this Comment we point out that the two theories make different
predictions, however, for the shape of $\Delta G(V)$, since
it is  so small ($\Delta G_{\rm max}<e^2/h$ for all samples)
that the parameters of {\em only one or 
two}\/ TLSs (labeled by $i=1,2$ below) seem to be modulated
by slow fluctuators. 
Since the TLS-electron couplings depend 
strongly only on the inter-well distance, but changes in the environment
mainly alter the well depths,  the parameters modulated most strongly
will be the TLSs' asymmetry energies $E_i$. Thus, 
 one should be able to fit $\Delta G(V)$ assuming induced 
telegraph fluctuations,  $E_i\Leftrightarrow E_i^\prime$,
for only a few TLSs. 

\begin{figure}
\centerline{\psfig{figure=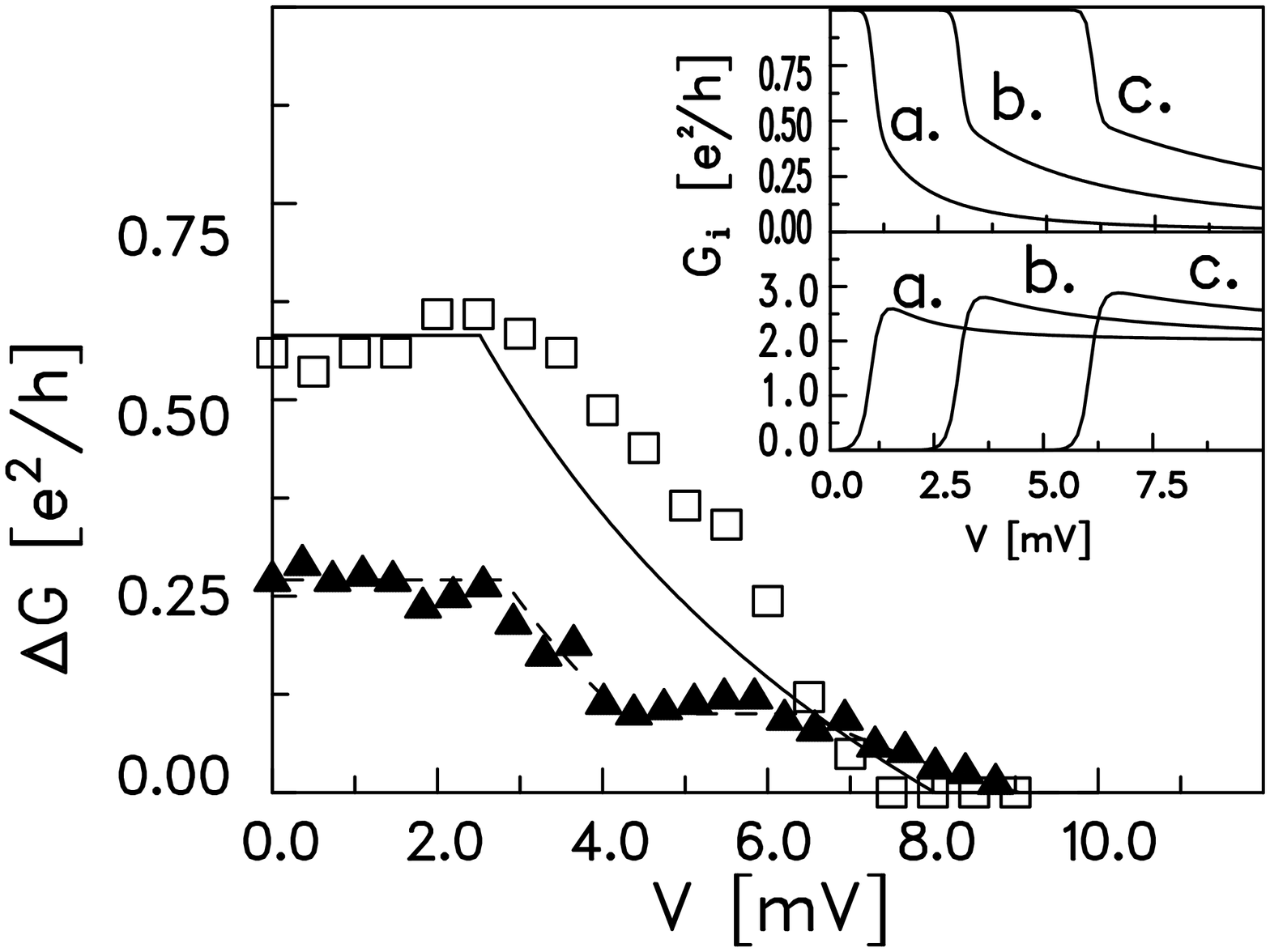,height=5cm}} 
\caption
{The squares give $\Delta G(V)$ for Fig.~2, curve 3 of \protect\cite{KSK}
and the triangles give the noise amplitude multiplied by 2 (for visibility)
of Fig.~4, curve 1 of \protect\cite{KSK} (uncertainties $\sim 0.1 e^2/h$). 
VZ's theory gives (a) the solid curve for $\Delta G(V)$ for $i=1$, 
with $E_1,E_1^\prime
\! = \! 8,3$meV, $\alpha_1 \!=\! 1$ and a Kondo temperature
$T^1_K$=17K; and (b) the 
dashed curve for $i=1,2$, with 
$E_{1,2} \! = \! 9,4.2$meV,
$E_{1,2}^\prime \! =\! 6.2,2.8$meV, $\alpha_{1,2} \! = \! 1,0.8$, 
and $T_K^{1,2} \! = \! 8.9, 6.2$K. 
The upper (lower) inset shows KK's \protect\cite{KK} 
(Kozub's \protect\cite{Kozub}) predictions for $G_i(V)$ for elastic 
(inelastic) scattering, for $E_i \! = \! 1,3,6$meV (a,b,c), at $T \! =
\! 1.2 K$. }
\end{figure}
\newpage

In VZ's scaling theory \cite{VZ},
the  renormalized (energy-dependent) dimensionless TLS-electron couplings become 
isotropic near the Kondo temperature $T_K^i$ \cite{VZ},
$v_i^{x,y,z} \! = \! v_i(\varepsilon)$, and  can be obtained
to leading logarithmic order 
by solving the scaling equation (4.5) of Ref.~\cite{VZ}(b);
since $E_i$ provides a lower cutoff for the scaling procedure,
$v_i (\varepsilon \! < \! E_i) \! \simeq \! v_i (E_i)$. 
Since  the corresponding scattering cross section  $\sigma_i (\varepsilon)$
is proportional \cite{VZ}(c) to $ k_F^{-2} v_i^2 (\varepsilon) $,
we estimate the ZBA contribution of TLS `$i$' 
as \cite{KK}  $G_i(V,E_i) \! \simeq \! - \alpha_i {2e^2\over h} 
v_i^2(eV)/v_{\rm fp}^2$, where 
$\alpha_i \! \simeq \! 1$ is a geometry-dependent constant \cite{vD},
and  we normalized  $G_i$ by the fixed point coupling $v_{\rm fp}$ 
to recover the unitarity limit $ 2e^2/h$ \cite{Ralph,vD}
at  $V \! = \!0$ if $ E_i \! = \!0$.
Thus each fast TLS is characterized by four parameters ($\alpha_i$,
$E_i$, $E_i^\prime$, $T_K^i$), and  
$\Delta G(V) = |\sum_i G_i(V,E_i) -  G_i(V,E_i^\prime)|$. 
Fig.~1 shows that the data for two samples of 
Ref.~\cite{KSK} can be fitted quite well 
using (a) one and (b) two TLSs, respectively. 

In contrast, the inset of Fig.1 shows KK's \cite{KK} prediction
for $G_i(V)$ for elastic scattering, and also Kozub's \cite{Kozub}
for inelastic scattering. Inspection shows that due to 
 these $G_i(V)$ curves' long (power-law) tails, 
it is impossible to fit $\Delta G(V)$ using
the difference $|G_i-G_i^\prime|$
(nor using a sum $|\sum_i (G_i-G_i^\prime)|$ for several TLSs):
$E_i\ll E_i^\prime$ gives too long a tail, and $E_i \simeq E_i^\prime$ 
too small a height for $\Delta G(0)$, even though, to obtain a maximally
large $G_i(0)\simeq e^2/h$, we took the TLS in the junction center (KK's 
$q=0.5$) and assumed extremely large effective cross-sections
($\simeq k_F^{-2}$).

In summary, KSK's experiments for the first time allow the measurement
of the conductance contributions of {\it individual} fast TLSs; the
$\Delta G(V)$ curves agree much better with VZ's than KK's theory. 
If both the $V$- and $T$-dependence of $\Delta G$ were known,
a $V/T$ scaling analysis \cite{Ralph,vD} could provide a further test for
VZ's scenario.

We  thank KSK  for discussions and sending us their data.
G.Z.\ was supported by the Magyary Scholarship and 
OTKA T021228, OTKA TO 24005/1997 
 and F016604, J.v.D.\ by ``SFB 195'' of the DFG.

\noindent
 G.\ Zar\'and,$^1$ Jan von Delft${}^2$  and 
A.\ Zawadowski${}^1$ \\
{\small
${}^1$Institute of Physics, Technical University of Budapest,\\
\hphantom{${}^1$}H 1521 Budafoki \'ut 8., Budapest, Hungary\\
${}^2$Institut f\"ur Theoretische Festk\"orperphysik, \\
\hphantom{${}^1$}Universit\"at Karlsruhe, 76128 Karlsruhe, Germany\\ 
\noindent
Received 23 May 1997}

\end{document}